\documentclass[
amsmath,amssymb]
{revtex4}

\usepackage{pstricks}
\usepackage{graphicx}
\usepackage{dcolumn}
\usepackage{bm}

\begin{document}

\newcommand{\locsection}[1]{\setcounter{equation}{0}\section{#1}}
\renewcommand{\theequation}{\thesection.\arabic{equation}}

\def\beqn{\begin{eqnarray}}
\def\eeqn{\end{eqnarray}}
\def\beq{\begin{equation}}
\def\eeq{\end{equation}}
\def\l{\left (}
\def\r{\right )}
\def\lq{\left [}
\def\rq{\right ]}
\def\fr{\frac}
\def\la{\label}
\def\hs{\hspace}
\def\vs{\vspace}
\def\inf{\infty}
\def\ran{\rangle}
\def\lan{\langle}
\def\ov{\overline}
\def\tl{\tilde}
\def\tm{\times}
\def\lrar{\leftrightarrow}





\vs{1cm}

\title{Neutrino Properties from $E_6 \times SO(3) \times Z(2)$}

\author{Berthold Stech}
\email{b.stech@thphys.uni-heidelberg.de}

\affiliation{Institut f\"ur Theoretische Physik, Philosophenweg 16, D-69120 Heidelberg, Germany}


\vs{1cm}

\date{February 18, 2010}

\begin{abstract}

The group $E_6$ for grand unification is combined with the generation symmetry group $ SO(3)_g \times Z(2) $. The coupling matrices
in the Yukawa interaction are identified with the vacuum expectation values of scalar flavons.  The symmetric part of this $3\times 3$ coupling matrix can be identified with the known $up$-quark hierarchy. The antisymmetric part is responsible for fermion mixings and $CP$ violation. 
Numerical fits with only few parameters reproduce quantitatively all known fermion properties. The model predicts an inverted neutrino hierarchy, CP violation properties as well as  the $0\nu 2\beta $ decay parameter. It also predicts that the masses of the two lightest of six `right handed' neutrinos lie in the low TeV region.

\end{abstract}

\maketitle

\vspace{-0.5cm}

\section{Introduction}\label{sec:1}

 Grand unified theories \cite{Pati:1974yy} provide an understanding of the structure and the quantum numbers of the standard model and relate all fermions of a given particle generation. The most symmetric and elegant unification symmetry is $E_6$ \cite{E6, eric,
 newE6, S+T:2008sb}:
 the standard model fermions are in the lowest representation ({\bf {27}}) of this group. 
Quark, antiquark and lepton fields occur only in singlet and triplet $SU(3)$ representations of the
maximal subgroup of $E_6$
\begin{equation}
\la{SU3}
 SU(3)_L  \times  SU(3)_R  \times SU(3)_C\equiv G_{333}~.
\eeq

Moreover, these fields transform into each other in a cyclic way. This presentation is based on the $E_6$ and flavor model of ref (\cite{S+T:2008sb}) but includes modifications.
 
Because of the existence of 3 generations any GUT symmetry should also unify the different generations and thus be extented to include a flavor symmetry.  In the simplest extention one uses the direct product GUT $\times $Flavor. We identify the  coupling matrices  in front of the Higgs fields as vacuum expectation values of new scalar fields (flavons), which are $E_6$ singlets but carry generation quantum numbers.
 
Using for the fermions left handed two component Weyl fields $\psi^\alpha$ (with $\alpha $ denoting the generations)  the Yukawa interaction is of the form 
\begin{equation}
\la{1.2}
  {\cal L}^{eff}_Y  =   \frac{\langle \Phi_{\alpha \beta} \rangle}{M} (\psi^{\alpha T} H \psi^{\beta})+ \cdots \\
 \eeq
     
  $\Phi_{\alpha \beta}$ describe the flavon fields, $  H $ stand for Higgs fields (two representations $\bf{27}$ and 
 the antisymmetric  representation $\bf{351}$  of $E_6$ are used) and $M$ gives the scale at which the effective Yukawa interaction of dimension 5 is formed. In (\ref{1.2}) $E_6$ indices are suppressed .

Clearly, Yukawa interactions of this form are effective ones and thus have to be understood on a deeper level. A renormalizeable model has been constructed by the introduction of additional heavy spinor fields \cite{S+T:2008sb}. However, in this presentation I will be concerned with the phenomenology of the effective Yukawa interaction only. 
For three generations 9 neutral flavon fields are needed. In $ \Phi_{\alpha \beta} $ they are combined to form a hermitian $3 \times 3 $ matrix. 

The introduction of the hermitian matrix field  $ \Phi (x) $ coupled to the fermion fields suggests to use the group $SO(3)_g$ to describe the generation symmetry. In addition, we will make use of a discrete parity-like symmetry, 
 $Z(2) =P_g$, generation parity. Thus, in our approach {\bf{all}} fermion fields are taken to transform as 3-vectors under the flavor group $SO(3)_g$. Consequently, the symmetric part of $\Phi_{\alpha \beta}$ has to transform as {\bf {1 + 5 }} and the antisymmetric part as {\bf {3}}. 
Spontaneous symmetry breaking of this symmetry leads to the vacuum expectation values of $ \Phi_{\alpha \beta} $ introduced above.
By an orthogonal transformation we can go to a basis in which the symmetric part of $\langle \Phi_{\alpha \beta} \rangle $ is diagonal. This part gives no flavor mixing but determines the fermion hierarchy. All flavor mixings and $CP $ violation arise from the antisymmetric part of the matrix $\langle \Phi_{\alpha \beta} \rangle $. Because of the Pauli principle only an antisymmetric Higgs representation (with respect to $E_6$ indices) can go together with this flavor changing coupling matrix. 

$ E_6$ unification cannot be obtained in a single step. It requires an intermediate symmetry. We take for it the group 
$G_{333}$ . Because of its $SU (3)$  content, this group can only be effective at and above the point where the gauge couplings 
$g_1$ and $g_2$ of the standard model meet, i.e. at the point of electroweak unification. Thus, the standard 
model itself fixes the mass scale of the intermediate symmetry. 
It is found to be $M_I \approx  2 \times 10^{13}~ GeV $.  
This mass is then also the scale for the masses of the heavy fermions occurring in the 27 representation. 
As in all GUT models one needs a small mixing of the standard model states with heavy states. In our $E_6$ model this mixing is described by a small $SU(3)_L$ and a small $SU(3)_R $ $U$-spin transformation of the antisymmetric Higgs field only. Thus, this mixing goes together with the flavor mixing by the antisymmetric part of the coupling matrix $\langle \Phi_{\alpha \beta} \rangle$. Since the $up$ quarks do not change under $U$-spin transformations they are not mixed. Thus, the experimentally observed $up$ quark hierarchy is the dominant ingredient for the hierarchical structure of all fermions.
Obviously, in our model there are strong relations between all fermions. The hierarchy, mixing angles and $CP$ properties in the quark and charged lepton sector influence to a large extent the corresponding hierarchy, mixing and $CP$ properties of all other fermions, in particular those of heavy and light neutrinos. In spite of these close connections and the very few parameters in the model, a quantitative fit to all known fermion properties can be achieved.

For notations and to have a look on the ${\bf{27}}$ representation needed below, this representation is shown here for a single fermion generation using the quantum numbers of the subgroup $G_{333}$

\begin{eqnarray}
\label{family}
\begin{array}{cc}
 & {\begin{array}{cc}
 \hspace{1.8cm}&
\end{array}}\\ \vspace{2mm}
(Q_L)_i^a=\hs{-0.2cm}
\begin{array}{c}
 \\
\end{array}\!\!\!\! &{\left(\begin{array}{ccc}
\hs{-0.3cm}u^a
\\
\hs{-0.2cm}d^a
\\
\hs{-0.2cm}D^a
\end{array}\hs{-0.2cm}\right)\! }~,\hspace{1.5cm}
\end{array}
\hs{-0.1cm}
\begin{array}{ccc}
& {\begin{array}{ccc}
 & &
\end{array}}\\ \vspace{2mm}
~~~~~~~~L^i_k= \hs{-0.5cm}
\begin{array}{c}
  \\
\end{array} \hspace{-0.1cm}&{\left(\begin{array}{ccc}
L^1_1 & E^{-} & e^{-}
\\
E^{+} & L^2_2 & \nu
\\
e^{+} & \hat{\nu } & L^3_3
\end{array}\right)~,}
\end{array}
~~~~~(Q_R)^k_a=\left( \hat{u}_a,~\hat{d}_a,~\hat{D}_a \right) ~. \hs{1.3cm}
\label {ac}
\eeqn
Here $i, k, a=1, 2, 3$.
In this description $SU(3)_L$ acts vertically (index $i$) and $SU(3)_R$
horizontally (index $k$) and  $a$ is a color index. 
It is seen, that in each generation there exist 12 new fields extending the standard model: colored quarks
$D, \hat{D}$, charged  leptons $E^+, E^-$ and 4 new neutral leptons $\hat{\nu }, L_3^3, L_1^1, L_2^2$,
which all must correspond to heavy, but not necessarily very heavy, particles.
A Higgs field in the ${\bf{27}} $ representation has the same form. For this field vacuum expectation values can occur at the 5 entries corresponding to the neutral places in $L^i_k$.

\section{The Effective Yukawa Interaction}\label{sec:2}

With respect to our symmetry
\begin{equation}
\label{2.1}
E_6 \times SO(3)_g \times P_g
\end{equation}
the fermion fields transform according to $ (27, 3, +)$  and the flavon fields $\Phi_{\alpha \beta}$ according to $( 1, 3 \times 3 , - )$. We need Higgs scalars: the $up$-quark Higgs  $H_u$ and the down quark Higgs $H_d$ transform as $(27, 1, - )$ and  $(27, 1, + )$, respectively. The antisymmetric Higgs scalar $H_A$ transforms according to $ (351, 1, - )$. $H_u$ is supposed to give masses to the $ up $ quarks and heavy states while  $H_d$ gives masses to the down quarks and heavy states of about the same strength. 
The difference between the top and bottom mass is supposed to arise due to the violation of $P_g$ symmetry which leads the fermions to couple to the combination $ H_{27} =cos z H_u + sin z H_d $ with $z$ fixing $m_b$. The orthogonal combination is denoted by $\tilde H_{27}$.

As mentioned in the introduction the  symmetric part of the matrix  $\lan \Phi_{\alpha \beta}\ran $ can be taken diagonal describing the $up$-quark hierarchy:
\begin{equation}
\label {G}
\hs{-0.8mm}G := \fr{\langle \Phi_{sym} \rangle }{M}= \hs{-0.8mm}\fr{1}{m_t}\hs{-0.8mm}\left(\begin{array}{lll}
m_u&0&0\\
0&m_c&0\\
0&0&m_t
\end{array}\right)\hs{-0.1cm}=\hs{-0.1cm}\left(\begin{array}{lll}
\sigma^4&0&0\\
0&\sigma^2&0\\
0&0&1
\end{array}\right)~.
\end{equation}
\nopagebreak
We took the mass ratios $m_c/m_t = m_u/m_c = \sigma^2$  to be valid at the high scale $M_I$. At the weak scale $ M_Z$  these ratios are modified. Taking $\sigma = 0.050$ gives good agreement with the experimental mass determinations. In the following we will use this parameter for expressing small quantities.
For the vacuum expectation value of the antisymmetric matrix  $\Phi_{antisym}$,  which should produce the mixings and CP violating properties of all fermions, we also suggest a simple form. 
It has a particular symmetry: it changes sign by exchanging the second with the third generation.
\begin{equation}
\label{A}
A: = \frac{1}{M'}\langle \Phi_{antisym}\rangle =
 {\rm i}~\left(
\begin{array}{ccc}
0&\sigma&-\sigma\\
-\sigma&0&1/2\\
\sigma&-1/2&0
\end{array}\right)~.
\end{equation}


Using the matrices (\ref{G}) and (\ref{A})  the effective Yukawa interaction can be written
\begin{eqnarray}
\label{Yukawa}
{\cal L}^{\rm eff}_Y~ = ~G_{\alpha \beta}
\left(\psi^{\alpha T} H_{27} \psi^{\beta} \right)+ A_{\alpha \beta}
\left(\psi^{\alpha T} H_{A}\psi^{\beta}\right)+
\frac{1}{M_N}
\left( G^2 \right)_{\alpha\beta}
\left((\psi^{\alpha T} H_{27}^{\dagger})_1(\tilde H_{27}^{\dagger} \psi^{\beta})_1 \right)~.\hs{0.9cm}~
\label{eff}
\eeqn

The first two terms in (\ref{Yukawa})  must clearly be there:  Vacuum expectation values of $H_{27}$ 
give masses to all light and heavy fermions, in particular also Dirac masses for the neutrinos. The mass hierarchy is universal and equal to the $up$-quark hierarchy. Vacuum expectation value of  $ H_A$ causes mixings in the down quark and charged lepton sector modifying thereby also the mass eigenvalues. However, the fields $\hat \nu $ and $L^3_3$ of eq.(\ref{family}) cannot get mass terms from the first two Yukawa interactions. A third interaction term is necessary. See reference  \cite{S+T:2008sb} for its suggested origin. In this third interaction term the fermion fields form singlets with the hermitian conjugate Higgs fields of the same representation and are multiplied with the flavon fields appropriate for these Higgses. We choose $ M_N = M_I $. Obviously, this third interaction give masses only to neutral leptons, providing them with the extremely strong flavor hierarchy $G^2$.
 
Replacing now the Higgs fields by their vacuum expectation values, the total mass matrix for the $3 \times 27 = 81 $ fermions can be written down:
\vspace{-0.13cm}
\begin{equation}
\label{M}
M^{r,r'}_{\alpha \beta}~ = ~G_{\alpha \beta}
\langle H_{27 }(r,r') \rangle+ A_{\alpha \beta}
\langle  H_{A}(r,r')\rangle  +
\frac{1}{M_I}
( G^2 )_{\alpha  \beta} 
\langle H_{27}^{\dagger~r} \rangle   \langle \tilde H_{27}^{\dagger~r'} \rangle .
\eeq

In this equation the $E_6$ indices $ r,r'$ run from $1$ to $27$. For simplicity, we left out the relevant Glebsch-Gordan coefficients. The mass matrix $ M^{r,r'}_{\alpha \beta} $
determines all fermion masses and mixings. In particular, after diagonalisation, it connects quark and charged lepton properties with the properties of light and heavy neutrinos. 
 $ \langle H_{27 } \rangle$ and $ \langle \tilde H_{27 } \rangle$  each have  5 possible places for vacuum expectation values, places which correspond to the neutral elements of the leptonic matrix in (\ref{ac}). However, one can use a basis such that $ \langle H_{27 } \rangle$ has only diagonal elements. They can be  identified with the masses (at the scale $M_I $) of the top quark, the bottom quark ($sin z \langle H_d \rangle^2_2 = m_b$) and the heavy down quark. For the latter we use the high mass scale available: $ m_D \simeq M_I $.
 In the basis chosen also the matrix $ \langle \tilde H_{27 } \rangle$ can be given a simple form: According to  (\ref{Yukawa}) only the high scale elements (situated in the last row) are important for this matrix.  
Thus we can set
\begin{equation}
\label{vevH}
\langle {H}_{27}\rangle    ~\simeq 
\left(\begin{array}{lll}
m_t &~~~0&~~~~0\\
~0& ~m_b &~~~~0\\
~0&~~~0&~ ~~m_D
\end{array}
\right)~~~,~~~~
 \noindent
\langle \tilde H_{27}\rangle  ~\simeq 
\left(\begin{array}{lll}
~0 &~~~0&~~~~0\\
~0& ~~cot~ z~ m_b&~~~~0\\
~0&~~m_D&~~\kappa~ m_D
\end{array}
\right)~.
\end{equation}

 Thus, the vacuum expectation values of both Higgs representations ${\bf{27}} $  are fully determined by $ m_t$, $m_b$ , $M_I$ and the constant $\kappa$ (expected to be $\lesssim1$) . For
 $m_t$ at the scale $M_I$ we will use $m_t(M_I)\simeq 131$ GeV. It corresponds to $g_t \simeq1$  for the coupling constant of the top quark at this scale. 
 
It remains to consider the vacuum expectation values of the antisymmetric Higgs representation $ H_A$. According to $E_6$ the relevant ones for quarks are in the matrix $f^i_k:= \langle H_A \rangle^i_k$  i.e. among the $ (3^* \times 3 ) $ elements with regard to $SU(3)_L \times SU(3)_R$ .  For the leptons the vacuum expectation values  occur in the $ (3^* \times 6^* )$ and the $ (6 \times 3 )$ elements: 
$f^{i,\{1,k\}} $ and $f_{\{1,i\},k}$~with~ $(i,k)\subset (2,3)$. We set $f^3_2 \approx f^{3\{1,3\}}$ , take two elements carrying the third $SU(3)$ index and  apply the small $U$-spin mixing angles $\theta_{L}$ and $\theta_{R} $  to account for the flavor mixing of high scale particles with the standard model particles. For instance, $ f^3_2 $ is given by $- \theta_{R}~f^3_3 $ and $f^2_2= \theta_L~ f^3_2$.
For the neutral leptons, discussed in detail below, one additional vacuum expectation value must be introduced. It is the standard model singlet  $f_A = f_{\{3,3\},1} = \langle H_A\rangle _{\{3,3\},1} $ which appears in the $ (6 \times 3 )$ representation of $ H_A $. It is part of a $ U_L$-spin triplet and thus expected to have a value very small compared to the ones for doublets. But being a standard model singlet it has nevertheless to be taken into account.

It is seen that even though our model describes 81 fermions with all their masses and mixings, the number of free parameters needed in (\ref{M}) is astonishingly small. We took all vacuum expectation values of Higgs fields to be real. Since the matrix $A$ is purely imaginary, flavor mixing leads to "maximal" $CP$-violation . 

\nopagebreak
\section{The down quark and charged lepton masses and mixings}\label{sec:3}

The two parameters $m_t$ and $\sigma$ describe the up quark masses very well. Thus we can immediately turn to the down quarks and charged leptons using (\ref{M}) and the vacuum expectation values introduced above.

The down quark mass matrix is shown as an example.
It includes the heavy $D$ quarks and is therefore  a $6\tm 6$
matrix. It has the form
\begin{equation}
\label{MdD}
\begin{array}{cc}
 & {\begin{array}{cc}
 \hspace{-0.1cm}\hat{d}\hspace{0.5cm}& \hspace{1.7cm}\hat{D}
\end{array}}\\
 \vspace{0.6cm}
M_{d, D}=
\begin{array}{c}
 d\\D
\end{array}\!\!\!\!\!\! &{\left(\begin{array}{ccc}
\hs{-0.1cm}m^0_b G+f_2^2 A~,&\hs{0.1cm} f_3^2 A
\\
\hs{-0.2cm}f_2^3A ~, &\hs{0.1cm}M_I G+f^3_3 A
\end{array} \right)\! }~.
\end{array}
\eeq
In this equation we denoted the input value  for the mass of the bottom quark by $m^0_b$ in order to distinguish it from the corresponding eigenvalue of the diagonalized matrix. The vacuum expectation values of the standard model singlet components of $H_A$, $f^3_3$ and $f^3_2$, turn out to have values small compared to $M_I$.  This allows to use  the see-saw formula to reduce the 
$ 6\times 6 $ mass matrix.  
Integrating out the heavy $D$-states gives us the $3\tm 3$ mass matrix for the down quarks of the standard model with only 3 real parameters: $m^0_b$ , $f^2_2$ and $f^2_3  f^3_2 / M_I $. They allow to fit  3  observables  and predict the remaining 4 out of seven :  3 mass eigenvalues, 3 mixing angles and the CP violating phase.

The result is very satisfactory as in \cite{S+T:2008sb}. The masses and the CKM matrix, in particular also the unitarity triangle, obtain values within experimental errors.



The charged lepton sector is constructed in a similar way. There remains some freedom to choose  specific values for the constants $f^3_2 \approx  
f^{3,\{1,3\}} $. However, due to the known mass of the vector boson $W$ of the standard model there is a relation between vacuum expectation values which has to be respected. It connects all constants which belong to $SU(2)_L$ doublets at the scale $M_I$ and thus gives an important restriction for $f^{3,\{1,3\}}$.  
It is useful to write
\begin{eqnarray}
\label{xL}
                     \frac{ f^{3,\{1,3\}}}{M_I}= \sigma^3 x_L ~, ~~  \frac{ f_A}{M_I} =\sigma^5 x_A ~.
 \eeqn
The power in $\sigma$   ensures that the  $3\times 3$ mass matrices are not becoming singular in the formal limit $\sigma \rightarrow 0$.  
The value $x_L\approx 0.04$ turns out to satisfy the requirements.  This fixes $cot  z~m_b \approx 97~GeV$ and 
gives consistent values for  $\theta_L $ and
$\theta_R$: ~$ \theta_L \approx - 2.2 \cdot 10^{-9} $~,~~$\theta_R \approx   0.0081$~.

\section{The Neutral Lepton Mass Matrix}\label{sec:4}

In each generation one has to deal with $5$ neutral leptons (see (\ref{ac})). Thus, the matrix for neutral leptons is a $15\tm 15$ matrix. Using (\ref{M}) it can be written as a $ 5 \times 5 $ matrix in which each entry stands for a $ 3 \times 3 $ flavor matrix. It is the matrix for the lepton fields $ \nu = L^2_3$ , the anti-lepton fields $ \hat \nu = L^3_2 $,  and the three fields $L^3_3 $, $L^1_1$ and $L^2_2$~.

\begin{equation}
\begin{array}{ccccc}
 & {\begin{array}{ccccc}
\hs{-0.2cm} L^2_3\hs{0.2cm} & \hs{1.8cm}L_2^3 \hs{1.8cm}
& \hs{1.8cm}L_3^3\hs{1.2cm} &\hs{2.2cm}L^1_1\hs{1.2cm} &
\hspace{2.4cm}L^2_2\hspace{0.2cm}
\end{array}}\\ \vspace{1mm}
M_L\hs{-0.1cm}=\hs{-0.1cm}
\begin{array}{c}
L^2_3 \\ L^3_2 \\ L^3_3 \\ L_1^1 \\ L^2_2
 \end{array}\!\!\!\!\! \hs{-0.1cm}&{\left(\begin{array}{ccccc}
~0~ & - m_t G  ~ &~ 0~ &~ -f^{3,\{1,3\}} A~ & ~0~
\\
-m_t G~   &~ 0 ~ &~ M_I G^2 +f_A A~ &~ 0~ &~ 0~
 \\
~0~ &~ M_I G^2+f_A A^T~ &~ \kappa M_I G^2 ~ &~ 0~ &~ m_t G~
\\
-f^{3,\{1,3\}} A^T~ &~ 0~ &~ 0~ &~ 0~  &~ M_I G + f^{3,\{1,2\}}A~
\\
~0~ &~ 0~ &~ m_t G~ &~  M_I G + f^{3,\{1,2\}}A^T~ & ~ 0~
\hs{-0.1cm}\end{array}\hs{-0.1cm}\right)}
\end{array}  \!\!
\label{ML}~.
\end{equation}
In the $12\tm 12$ sub-matrix for the heavy leptons we neglected small terms like $f^2_2$ and $f_{\{1,3\},3}$. They play no role in the evaluation of the light neutrino properties. The flavor matrices $G$ and $A$ appear as dictated by the general mass matrix (\ref{M}). There are two parameters not yet fixed: $\kappa$ and $f_A$.

In order to study the submatrix relevant for the light neutrinos it is useful to rewrite the matrix (\ref{ML}) in the form
\begin{equation}
\begin{array}{ccc}
 & {\begin{array}{ccc}
\hspace{0.2cm} \hspace{0.7cm} &
&
\end{array}}\\ \vspace{1mm}
{M_L}=\hs{-0.2cm}
\begin{array}{c}
 \\  \\
 \end{array}\!\!\!\!\!\hs{-0.2cm} &{\left(\begin{array}{cc}
0 &\Omega
\\
\Omega^T   &\hat{M}
\end{array}\right)}~,
\end{array}  \!\!  ~~~~~
\label{neu-ML}
\end{equation}
where $0$ stands for the $3\tm 3$ zero block matrix, while $\Omega $ and $\hat{M}$ are $3\tm 12$ and $12\tm 12$ matrices, respectively.

The matrix $\hat{M}$ contains the masses of the heavy neutral states which should be integrated out. Neglecting~$f^{3,\{1,2\}}$ which is small compared to $M_I$ the light neutrino $3\tm 3$ mass matrix can now be obtained via the see-saw formula
\begin{equation}
m_{\nu }~ \simeq -\Omega \hat{M}^{-1}\Omega^T~\simeq ~~ \fr{(m_t)^2}{M_I} 
 \left(  \kappa ~ {\bf{1}}+\frac{ f^{3,\{1,3\}}}{M_I}(A\fr{1}{G}+\fr{1}{G}A^T)+ \cdots \right)
\la{mnu}
\eeq

We use ( \ref{xL}) and the further abbreviation $m= m_t^2/M_I$. Keeping then only terms in $\sigma$ and $x_A$ up to first order the light neutrino mass matrix takes the very simple form

\begin{equation}
\begin{array}{ccc}
 m_{\nu}\hs{0cm}=m~ x_L \hs{-0cm}
{\left(\begin{array}{ccc}
\vspace{0.3cm}
~\frac{\kappa}{x_L}~ & - i~ &~ i~  \\

\vspace{0.3cm}

- i~   &~ \frac{\kappa}{x_L} -2 x_A~ &~ -i\frac{\sigma}{2} +x_A~ \\
 
~i~ &~ -i\frac{\sigma}{2}+x_A~ &~ \frac{\kappa}{x_L} ~
\hs{-0.1cm}\end{array}\hs{-0.1cm}\right)}
\end{array}  \!\!
\label{mnu}~.
\end{equation}

(For a detailed investigation a correction to the see-saw formula has to be taken into account, see 
\cite{S+T:2008sb}).

The mass squared eigenvalues are obtained from
$m_{\nu} \cdot m^{\dagger}_{\nu}$. In the limit $x_A \to 0$  they are to first order in $\sigma $ given by
\begin{eqnarray}
(m_2)^2 \simeq (\kappa^2 +x_L^2 ~(2 + \fr{\sigma}{\sqrt{2}}))~ m^2~,
\nonumber
\\
(m_1)^2 \simeq (\kappa^2 +x_L^2~(2 - \fr{\sigma}{\sqrt{2}}))~ m^2~,
\nonumber
\\
(m_3)^2 \simeq \kappa^2 m^2~.\hs{1cm}~
\label{eival}
\eeqn

It is now easy to see the following properties of the neutrinos:

{\bf i)}  The mass spectrum for the light neutrinos has the inverted form.

{\bf ii)} In the no mixing limit $x_L \to 0$  the 3 neutrino masses are degenerate.

{\bf iii)} The ratio between the solar and atmospheric mass squared differences is
$\fr{\sigma }{\sqrt{2}}\simeq  0.035$ independent of the parameters  in good agreement with experiments.

{\bf iv)} The experimentally observed atmospheric mass squared difference can be used to fix the parameter
 $  x_L $ :
\begin{equation}
\label{At}
 x_L \simeq \frac{1}{\sqrt{2} m} \sqrt{\Delta m^2_{\rm atm}} \simeq 0.038~.
\eeq
This value of $x_L$ is consistent with the requirement for this quantity as discussed in the section for charged leptons.

{\bf v)} In the limit ~$x_A\to 0~ (f_A \to 0)$ , the neutrino mixing matrix is of {\it bimaximal} form.

  {\bf vi)} In (\ref{At})~ $m$ depends on our starting value $M_I\simeq 2 \cdot 10^{13 }~GeV$. Thus, the heaviest of the heavy neutrinos has about the same mass value as seen from (\ref{neu-ML}). 
  
Remarkably,
the spectrum of the heavy neutrinos is a super hierarchical one, determined by the square of the $up$ quark mass spectrum ($G^2$). Thus, the two lightest ones of the heavy neutrinos have masses $ \simeq \sigma^8~M_I $ i.e. 
their masses are in the $ \lesssim  TeV$ region.

In \cite{S+T:2008sb}, by preferring values of $\kappa \lesssim x_L$, a detailed calculation of the light neutrino properties could be performed. The inclusion of the constant $x_A$ and of renormalization effects accounted for the experimentally observed deviations from {\it bimaximal} mixing. All neutrino properties including the $CP$ violating phase have been predicted for this case. 

Higher values of the neutrino masses are also of great interest, in particular, in view of the ongoing KATRIN \cite{Katrin} and GERDA \cite{Gerda} experiments. In our model the mass of the lightest neutrino is given by $ \simeq 0.858 ~\kappa~ eV $. The remaining parameter $x_A$ is then used for the fit to the known neutrino data (essentially the solar mixing angle).  In this preliminary investigation
we are using effective values for $\kappa$ and $x_A$ and ignore renormalization group effects. The numerical diagonalization of (\ref{ML}) gives the 15 eigenvalues for the masses and the full $15 \times 15 $ mixing matrix. The $ 3 \times 3$ sub-matrix $U$ we are mainly interested in is not strictly unitary. The largest deviations occur in the $(2,3)$ and $(3,3)$ elements of $U$ with a magnitude of about  $7 \cdot 10^{-4} $. 

As an example for large neutrino masses leading to an almost degenerate spectrum we take  $\kappa =0.30 $ and $x_A= - 0.0121 $. One obtains for the lightest neutrino the mass $ 0.257~ eV $.
The  mass square differences obtained are
\begin{eqnarray}
\Delta m_{\rm sol}^2=m_2^2-m_1^2 \simeq 7.7 \cdot 10^{-5}~{\rm eV}^2~,~
\Delta m_{\rm atm}^2 =m_2^2-m_3^2 \simeq 2.2 \cdot 10^{-3}~{\rm eV}^2~.
\la{delmass}
\eeqn
For the neutrino mixing angles, and the CP violating phase $\delta_l$ (appearing in the neutrino oscillation amplitude)  one finds after including the effect of the diagonalization of the charged lepton mass matrix

\begin{eqnarray}
\theta_{12}^{\nu} \simeq 32^o~(32^o)~,~~~\theta_{23}^{\nu} \simeq 44^o~(40^o)~,~~~\theta_{13}^{\nu} \simeq 3.0^o~(0.6^o)~,~~\delta  \simeq 82^o~ (264^o)~.
\la{angles1}
\eeqn
The values in parenthesis are the ones before diagonalization of the charged lepton mass matrix.
The mixing angles together with the mass squared differences (\ref{delmass}) are in good agreement with
global fits to neutrino oscillation data \cite{Fogli}. 
For the sum of the three masses ($ \Sigma $), the mass observable at $\beta$ decay ($ m_{\beta}$) and for the neutrino less double $\beta $-decay parameter ($ m0_{\beta \beta} $) one finds for $\kappa = 0.30 $
\begin{eqnarray}
\la{beta}
 \Sigma = m_2 +m_1+ m_3 \simeq ~0.78 ~eV, \nonumber \\
 m_{\beta} = \sqrt {|U_{11}|^2 m_1^2+|U_{12}|^2 m_2^2+|U_{13}|^2 m_3^2} \simeq ~0.26~eV~, \nonumber \\ 
 m_{0 \beta \beta} = |U_{11}^2 m_1+U_{12}^2 m_2+U_{13}^2 m_3 |  \simeq 0.25~eV~.
\eeqn

 We note that oncee $\kappa$ is specified there is not much freedom within this $E_6$ model to get different 
 oscillation parameters,  or to change the neutrino hierarchy. Remarkably, (\ref{beta}) holds to a good approximation also for different values of $\kappa$. One simply has to scale the right hand sides according to $\kappa/0.30$. 

$\nu_{\mu}$ and $\nu_{\tau} $ mix to about $3\%$ with the first generations of the neutral leptons
$L^1_1$ and  $L^2_2$ . $\nu_e$ mixes to about $10^{-6}$ with the lightest of the heavy neutrinos (a combination of the first generations of $\hat \nu=L^3_2$ and $L^3_3$).
\vspace{0.4cm}

Our phenomenological treatment gives a clear picture of the spectrum and mixings of the standard model fermions and of  their heavy $E_6$ partners. A quantitative fit reproducing all known properties of the standard model  charged and neutral fermions could be performed. Several not yet measured quantities are predicted. However, the model is still incomplete, because an understanding of the scalar sector, in particular of the vacuum expectation values of the $E_6$ Higgs fields, is not yet achieved.

\vs{0.4cm}

\hs{-0.5cm}
{\bf Acknowledgments}
\vs{0.1cm}

\hs{-0.5cm}

I like to thank Zurab Tavartkiladze for our very enjoyable collaboration and I also like to express my thanks to the organizers of the pleasant and fruitful meeting in Corfu.

\bibliographystyle{unsrt}

\begin{thebibliography}{99}


\bibitem{Pati:1974yy}
  J.~C.~Pati and A.~Salam,
  Phys.\ Rev.\  D {\bf 10} (1974) 275;
H.~Georgi and S.~L.~Glashow,
  Phys.\ Rev.\ Lett.\  {\bf 32} (1974) 438.


\bibitem{E6}
F.~Gursey, P.~Ramond and P.~Sikivie,
  Phys.\ Lett.\  B {\bf 60} (1976) 177;
Y.~Achiman and B.~Stech,
  Phys.\ Lett.\  B {\bf 77} (1978) 389;
Q.~Shafi,
  Phys.\ Lett.\  B {\bf 79} (1978) 301;
R.~Barbieri, D.~V.~Nanopoulos and A.~Masiero,
  Phys.\ Lett.\  B {\bf 104} (1981) 194.



\bibitem{eric}
For an early review see
B. Stech, Ettore Majorana Inst. Sci., Vol. 7 (1980) p. 23,
eds. S. Ferrara et al.


\bibitem{newE6}
Y.~Achiman  and A.~Lukas,
  Nucl.\ Phys.\  B {\bf 384} (1992) 78;
J.~L.~Rosner,
  Phys.\ Rev.\  D {\bf 61} (2000) 097303;
J.~D.~Bjorken, S.~Pakvasa and S.~F.~Tuan,
  Phys.\ Rev.\  D {\bf 66} (2002) 053008;
M.~Bando and T.~Kugo,
  Prog.\ Theor.\ Phys.\  {\bf 101} (1999) 1313;
 {\it ibid} {\bf 109} (2003) 87;
N.~Maekawa and T.~Yamashita,
  Prog.\ Theor.\ Phys.\  {\bf 107} (2002) 1201;
J.~Harada,
  JHEP {\bf 0304} (2003) 011.
  B.~Stech and Z.~Tavartkiladze,
  Phys.\ Rev.\  D {\bf 70} (2004) 035002.
 C.~R.~Das and L.~Laperashvili,
  Phys.\ Rev.\  D {\bf 74} (2006) 035007;
 J.~Sayre, S.~Wiesenfeldt and S.~Willenbrock,
  Phys.\ Rev.\  D {\bf 75} (2007) 037702;
 F.~Caravaglios and S.~Morisi,
  Int.\ J.\ Mod.\ Phys.\  A {\bf 22} (2007) 2469.
 S.F. King, S. Moretti, R. Nevzrov, Phys.Rev. D {\bf73} (2006) 035009~.
 
\bibitem{S+T:2008sb}
  B.~Stech and Z.~Tavartkiladze,
  ``Genertion Symmetry and E6 unification''
  Phys.\ Rev.\  D {\bf 77} (2008) 076009.
  [hep-ph/0311161].


\bibitem{Katrin}
A.Osipowicz et al. [KATRIN Collaboration], arXivhep-ex/0109033.

\bibitem{Gerda}
[GERDA Collaboration] Nucl.Phys.Proc.Suppl. 188:68-70, 2009.




\bibitem{Fogli}
G.L. Fogli, E. Lisi, A. Marrone, A.Palazzo and A.M. Rotunno, Phys. Rev. Lett. {\bf 101},
141801 (2008).

\end{thebibliography}

 \end{document}